\documentclass[preprint,12pt]{elsarticle}
\usepackage{amssymb}
\usepackage{epsfig,graphicx,bm}
\usepackage{natbib}
\journal{Icarus}
\begin{document}
\begin{frontmatter}
\title{The dynamical evolution of escaped Jupiter Trojan asteroids, link to other minor body populations}
\author[1,2]{Romina P. Di Sisto}
\ead{romina@fcaglp.unlp.edu.ar}
\author[3]{Ximena S.  Ramos}
\author[4]{Tabar\'e Gallardo}

\address[1]{Facultad de Ciencias Astron\'omicas y Geof\'\i sicas Universidad Nacional de La Plata}
\address[2]{Instituto de Astrof\'{\i}sica de La Plata, CCT La Plata-CONICET-UNLP  Paseo del Bosque 
S/N (1900), La Plata, Argentina}
\address[3]{Instituto de Astronom\'{\i}a Te\'orica y Experimental (IATE), Observatorio Astron\'omico, Universidad Nacional de C\'ordoba, Laprida 854, X5000BGR C\'ordoba, Argentina }
\address[4] {Departamento de Astronom\'{\i}a, Facultad de Ciencias, Universidad de la Rep\'ublica, Igu\'a 4225, 11400 Montevideo, Uruguay}

\begin{abstract}

The Jupiter Trojans  constitute an important asteroidal population both in number and also in relation to their dynamical 
and physical properties. They are asteroids located around $L_4$ and $L_5$ Lagrangian points on relatively stable orbits, in $1:1$ 
mean motion resonance with Jupiter. However, not all of them lie in orbits that remain stable over the age of the Solar System. 
Unstable zones allow some Trojans to escape in time scales shorter than the Solar System age. This may contribute to populate other 
small body populations. In this paper, we study this process by performing long-term numerical simulations of the observed Trojans, focusing on the 
trajectories of those that leave the resonance. The orbits of current Trojan asteroids are taken as initial conditions 
and their evolution is followed under the gravitational action of the Sun and the planets. 
 We built ``occupancy maps'' that represent the zones in the Solar System where escaped Trojans should  be found.
We find the rate of escape of Trojans from $L_5$,  $\sim 1.1$ times greater than from $L_4$. The majority of escaped Trojans 
have  encounters with Jupiter although they have encounters with the other planets too.
The median lifetime of escaped Trojans in the Solar System is $\sim 264000$ years for $L_4$ and $\sim 249000$ years for $L_5$.
 Almost all escaped Trojans reach the comet zone, $\sim 90 \%$ cross the Centaur zone and only $L_4$ Trojans
reach the transneptunian zone. Considering the real asymmetry between $L_4$ and $L_5$, we show that 18 $L_4$ Trojans and 14 $L_5$ Trojans with diameter $D > 1$ km 
are ejected from the resonance every Myr.  The contribution of the escaped Trojans to other minor body populations would be negligible, 
being the contribution from $L_4$ and $L_5$ to Jupiter-family comets (JFCs) and no-JFCs almost the same, and the $L_4$ contribution to 
Centaurs and TNOs, orders of magnitude greater than that of $L_5$. Considering the collisional removal, besides the dynamical one, 
and assuming that Trojans that escape due to collisions follow the same dynamical behavior that the ones removed by dynamics, we would 
have a minor contribution of Trojans to comets and Centaurs. However, there would be some specific regions were escaped Trojans could be
important such as Asteroids in Cometary Orbits (ACOs), Encke-type comets, Shoemaker-Levy 9-type impacts on Jupiter and Near-Earth objects (NEOs).

\end{abstract}

\end{frontmatter}
     
\section{Introduction} 
\label{intro}

 Jupiter Trojans are a population of asteroids in $1:1$ mean motion resonance (MMR) with Jupiter and are located within the $L_4$ 
and $L_5$ Lagrange points. It is a significant asteroidal population, both in number and also in relation to their dynamical 
and physical properties. 
They form a  key population for revealing the history of the Solar System since its existence and survival constrains the 
theories of formation and evolution of the Solar System as a whole (e.g. Marzari and Scholl, 1998; Morbidelli et al., 2005; 
Nesvorn\'y et al., 2013). 

 A distinctive feature of this population is an asymmetry in the number of bodies of the leading and trailing clouds. It is observed 
that the number of Trojans in $L_4$ doubles the number in $L_5$.  However, dynamical studies of the Trojan region show that both $L_4$ 
and $L_5$ have the same structure and stability. It is unclear whether the observations could be biased. For example, Grav et al. (2011) 
estimate that the real asymmetry between $L_4$ and $L_5$ should be corrected to a factor $1.4$. Nevertheless, even after correcting 
for biases, this asymmetry seems to be real and deserves attention.

The general dynamical properties of Trojans are also key to understanding the Solar System process. They
have been broadly studied in the past, both analytically and numerically  (e.g.  Erdi, 1996; Mikkola and Innanen, 1992; Milani, 1993). 
The first long-term numerical integration of Jupiter Trojans was made by Levison et al. (1997). They 
numerically integrated the orbits of 270 fictitious $L_4$ Trojans for 1 Gyr and also the orbits of 36 real Trojans for 
4 Gyr detecting stability areas and the places occupied by the real Trojans. They also followed the evolution of the 
escaped Trojans and studied their relation with Jupiter-family comets (JFCs).  A semi-analytical model to describe the long-term 
motion was developed by Beaug\'e and Roig (2001) where they identified and confirmed the existence of the majority of 
the families previously detected.

Thanks to the increasing computing power, long-term numerical simulations with a larger number of particles are now possible. 
 Recently, new long-term numerical integrations were carried out for the Trojans, which allowed us to deeper characterize the dynamics 
within the resonance. For example, Marzari and Scholl (2000, 2002) and Marzari et al. (2003) performed a series of numerical 
simulations to study the role of secular resonances in the dynamical evolution of Trojans and explored their stability properties 
and destabilization mechanisms. They found that direct perturbations made by Saturn are the main source of instability on time scales 
of the order of $10^7-10^8 $ years, while secular resonances, in particular, $\nu_{16}$ contribute on longer timescales.
This secular resonance raises the inclinations up to values greater than $20^{\circ}$ on a time scale of $10^8$ years. 
More recently, Robutel et al. (2005) and Robutel and Gabern (2006) performed long-term numerical simulations to study the global 
dynamical structure of the $L_4$ region.  They found that the inherent instability of the Trojans appears purely gravitational and 
caused by secondary and secular resonances within the tadpole regions.

Tsiganis et al. (2005) studied the stability of Trojans to define the effective stability of the region and compare 
it with the real distribution of Trojans. The effective stability region is defined in terms of the Lyapunov time and the escape time 
(time for an encounter with Jupiter) to study the regular and chaotic orbits at the border of the stability region. 
These orbits remain for the age of the Solar System. These authors numerically integrated real and fictitious $L_4$ Trojans finding 
that $17 \%$  of the real (numbered) Trojans escaped from the swarm over the age of the Solar System and that chaotic diffusion is 
the origin of the unstable population.

 Studies of the physical properties of Trojan asteroids suggest that they contain water ice and organic material, similar to the cometary 
  nuclei. Spectroscopic studies derive mainly D taxonomic classes but also some P and C classes (Fornasier et al., 2007; Emery et al., 2011; 
  Grav et al., 2012). 
  Water ice content and their similarity with cometary nuclei seems to indicate that Trojans could be formed in the 
outer Solar System. However, Emery et al. (2011)  found two spectral groups which they attributed to different intrinsic compositions 
and suggested two distinct regions of origin. There are several studies of albedos that show values in the range from 0.025 to 0.2 
(e.g  Grav et al., 2011; Grav et al., 2012; Fern{\'a}ndez et al., 2003) with a possible correlation between albedo and size, presenting lower 
values for smaller Trojans (Fern{\'a}ndez et al., 2009).  There are two space missions that plan to study in detail the physical 
properties of Trojans that will radically improve what we know about Trojans. ``Lucy'' NASA mission, which will be launched in 2021, 
 will encounter one Main Asteroid Belt and six Trojans from both swarms after a 12-year journey. Its main objective is to study the 
Trojan surface compositions, the diversity of taxonomic classes and also the interior and bulk properties to link those results with 
the source Trojan regions. The Japan Aerospace Exploration Agency, JAXA, is planning a mission to Trojans based on propulsion by 
a solar power sail (already tested by IKAROS, the first deep space solar sail). They plan to arrive on a Trojan target for global 
remote observation, surface and sub-surface sampling by a lander, and a possible sample return option.

Although Trojan asteroids are librating about the Jupiter's $L_4$ and $L_5$ stable equilibrium points, there exist regions of partial 
instability from which Trojans can escape from the resonance (Levison et al., 1997; Di  Sisto et al., 2014). In particular,
Di  Sisto et al. (2014), hereinafter D14, studied the dynamical evolution of Jupiter Trojans by numerical 
integrations of observed Trojans under the gravitational action of the Sun and the four giant planets. They focused 
their study on the properties of the observed population and the escape/survive population. They found that the escape 
rate of $L_5$ Trojans is greater than that of $L_4$,  and this fact could be responsible for $\sim$10 $\%$ of the total asymmetry. 
We will discuss those results later.
 
 In this paper, through numerical simulations,  we study the escape of Trojan asteroids and follow their dynamical evolution 
over the age of the Solar System or until they physically collide with a planet or they completely escape the Solar System. 
The main objective is to evaluate the dynamical routes of escape and the contribution 
of Trojans to other minor body populations, such as Comets, Centaurs, and NEOs. Also, we show the temporary captures of escaped 
Trojans in MMR, both inside and outside Jupiter's orbit.
 The paper is organized as follows. 
In Section 2, we review the main physical and dynamical characteristics of the Trojans, focusing on their number and 
size distribution. In Section 3, we describe the numerical simulations, while in Section 4, we present the 
results of a long-term integration of the observed Trojans. The Trojan contribution to other small body populations 
is shown in Section 5. We finally present our discussion and conclusion in Section 6.

\section{The Observed Population}

\subsection{Physical properties and size distribution}

Trojan asteroid spectral features are similar to those of cometary nuclei. Spectroscopic studies derive mainly  D 
taxonomic classes and some P and C (Fornasier et al., 2007; Emery et al., 2011; Grav et al., 2012). 
Grav et al. (2011) derived thermal models for 1739 Jovian Trojans, observed by the WISE survey (Mainzer et al., 2011), and
detected no differences for the leading and trailing cloud. They also found that the size distributions of the two swarms are 
very similar. Later,  by recomputing thermal model fits derived from that sample, Grav et al. (2012) calculated visible albedos 
that vary from 0.025 to 0.2 for small Trojans, with a median value of 0.05 for $D > 30$ km and  0.07 for $D<30$ km. In this paper we 
will adopt  those results from Grav et al. (2011), Grav et al. (2012) since they are based on the largest sample of 
albedo measurements.  

The size distribution of Jovian Trojans has been studied from observational surveys and measurements of 
albedos. Jewitt et al. (2000) carried out a survey in the $L_4$ direction, detecting $93$ 
Trojans with diameters of 4 km $ < D < $ 40  km, and obtained a Trojan cumulative size distribution 
(CSD) as a power law given by $N(>D) \propto D^{-s}$ with an index  $s=2.0\pm 0.3$.  But, by adding  
cataloged Trojans to the sample, they inferred  that there must be a break in the CSD at diameters $D \sim 80$ km toward
 an index  $s=4.5$.
New surveys for small Trojans performed by Yoshida and Nakamura  (2005), Yoshida and Nakamura  (2008) found that the faint end of the CSD seems 
to have another break around $D \sim 4-5$ km.

Szab{\'o} et al.  (2007) analyzed the observations of more than 1000 Trojans and found that the CSD of $L_4$ and $L_5$ are 
virtually the same with a cumulative index $s = 2.2$ in the range $10 $ km $\lesssim D \lesssim  80$ km, but there 
are $1.6 \pm 0.1$ more objects in the leading swarm than in the trailing one. 
A new and complete analysis of the magnitude distribution of $L_4$ Trojans has been recently done by Wong and Brown  (2015).  
From a Subaru survey, they detected 557 small $L_4$ Trojans, and, by  combining these observations with the bright Trojans contained 
in the MPC catalog, they fit a complete magnitude distribution in the range $7.2 < H < 16.4$ given by: 
\begin{equation}
  \label{md}
  \Sigma (H) = 
  \left \{
  \begin{array} {lr} 
   10^{\alpha_0(H-H_0)}, \qquad \qquad \qquad \qquad\qquad  H_0 \leq H \leq H_{b'}  \\
   10^{\alpha_1(H-H_{b'})} 10^{\alpha_0(Ḥ_{b'}-H_0)} ,  \qquad \qquad H_{b'} \leq H \leq H_b    \\ 
   10^{\alpha_2(H-H_{b})} 10^{\alpha_1(Ḥ_{b}-H_{b'})} 10^{\alpha_0 (Ḥ_{b'}-H_{0})} , \qquad H \geq H_{b}
  \end{array}
  \right. 
\end {equation}

The magnitude distribution begins at $H_0 = 7.22$ and has two breaks at magnitudes $H_{b'}=8.46 $ and $H_{b}=14.93 $.  
In the three regions defined by those limiting magnitudes the slopes are $\alpha_0 = 0.91$, $\alpha_1 = 0.44$ and 
$ \alpha_2 = 0.36$  

\begin{figure}[t!]
  \centerline{\includegraphics*[width=0.9\textwidth]{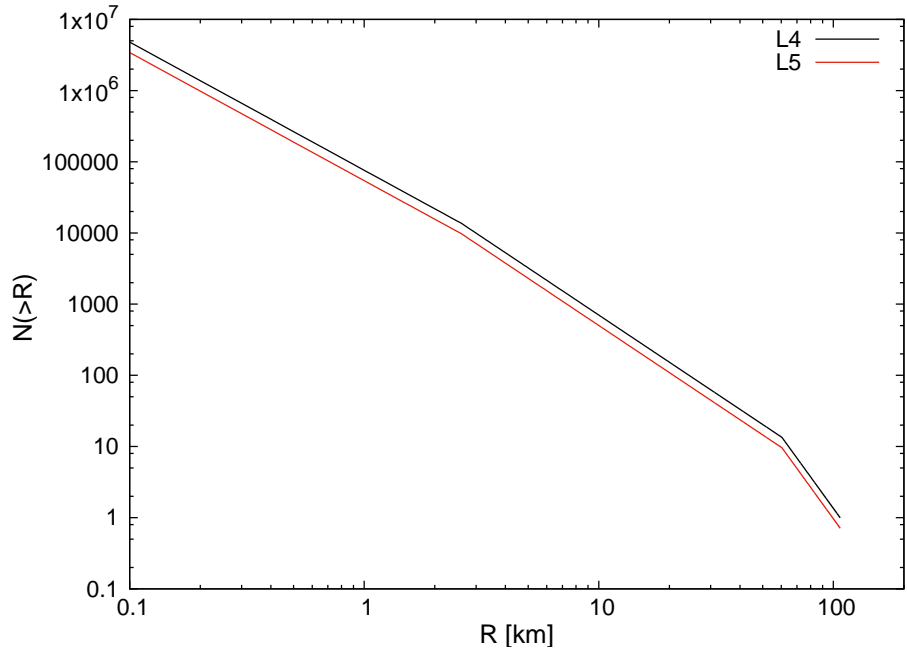}}
  \caption{Cumulative size distribution of $L_4$ (black) and $L_5$ (red) Trojans.}
  \label{csdfig}
\end{figure}

From a power-law magnitude distribution of the form  of Eq. (\ref{md}), the radii of Trojans follow a 
differential size distribution (DSD) given by $N(R)dR = C R^{-q} dr$, where $q = 5 \alpha + 1$, and $C$ is a constant. 
By converting Eq. (\ref{md})  
to the DSD in each region and integrating them, we can obtain the CSD of $L_4$ Trojans as:
\begin{equation}
  \label{csd}
  N(>R) = 
  \left \{
  \begin{array} {cc}
    (R/R_0)^{1-q_0} , \qquad \qquad \qquad \qquad\qquad\qquad  R_{b'} \leq R \leq R_{0}  \\
    (R/R_{b'})^{1-q_1}  (R_{b'}/R_0)^{1-q_0}    , \qquad \qquad \qquad   R_{b} \leq R \leq R_{b'}  \\ 
    (R/R_{b})^{1-q_2}  (R_b/R_{b'})^{1-q_1} (R_{b'}/R_0)^{1-q_0}     , \qquad\qquad R \leq R_{b}  
  \end{array}  \right. 
\end{equation}

where $q_0=5.55$, $q_1=3.2$ and $q_2=2.8$. To convert magnitude to radius, we consider the results 
of Grav et al. (2012) that obtained an albedo equal to 0.05 for $D > 30$ km  and  0.07 for $D<30$ km. So, the limiting 
radii are $R_0 = 106.9$ km, $R_{b'} = 60.4$ km and $R_{b} = 2.6$ km,  

We can see that the indexes of the CSD in the different ranges of radius, and also the location of the breaks, 
are in agreement with the previous studies. Then, we will adopt Eq. (\ref{csd}) as the CSD of $L_4$ Trojans. 
There is no comprehensive study on the CSD of $L_5$ Trojans, but some studies obtained very small (or no) differences 
in the CSD of $L_4$ and $L_5$ (Szab{\'o} et al., 2007; Yoshida and Nakamura, 2008). Then, based on the study 
of Grav et al. (2011),  we will consider that the number of $L_4$ Trojans is $1.4$ the number of $L_5$ Trojans, and 
that this asymmetry does not depend on the size. Therefore, the CSD of $L_5$ Trojans is given by Eq. (\ref{csd}) but 
offset by the asymmetry factor.  Both CSDs are plotted in Fig. \ref{csdfig}. 
Then, for example, there would be $ \sim  265000$ $L_4$ Trojans and $ \sim 190000$ $L_5$ Trojans with diameter greater than 1 km.

\subsection{Dynamical properties}

An analysis of the observed population was made by Di  Sisto et al. (2014), who found some differences in 
$L_4$ and $L_5$ swarms. While the mean values of the semimajor axis and eccentricities are almost the same for $L_4$ and $L_5$, 
the mean inclination in $L_5$ is $4^{\circ}$ greater than that of $L_4$. The relatively more excitation of the $L_5$ population 
is also appreciated in the distribution of inclinations, which is broader than that of $L_4$. 
Another point addressed by D14 what is related to the observed population is the calculation of proper elements 
and the determination of family members. They worked with numbered and multioppositional Trojans and concluded that 
only numbered asteroids have sufficiently well determined orbits to allow for detailed and long-term dynamical analysis.

\section{The Numerical Simulation}

The initial conditions of our simulations are the orbits of all numbered Jupiter Trojan asteroids as of March
 2013. Thereby, a numerical integration of 1975 $L_4$ Trojans and 997 $L_5$ Trojans were performed under the gravitational 
influence of the Sun and the planets from Venus to Neptune with the hybrid integrator EVORB (Fern{\'a}ndez et al., 2002). The time 
step was set to 7.3 days, which is roughly $1/30$ of Venus orbital period, and each Trojan evolved for 4.5 Gyr, 
unless removed due to a collision with a planet or the Sun, or due to reaching a heliocentric distance 
$r > 1000$ au. The encounters at less than 2.1 Hill's Radius with the planets were registered  to 
analyze them and to define an ``escape time'' for each ``escaped Trojan''. If a Trojan has an encounter 
with a planet, usually Jupiter, the time of the first encounter is considered the ``escape time'' and the Trojan would be 
an ``escaped Trojan''; its subsequent evolution through the Solar System up to a collision or escape will be 
the objective of this paper. 
    
\begin{figure}[t!]
  \centerline{\includegraphics*[width=0.9\textwidth]{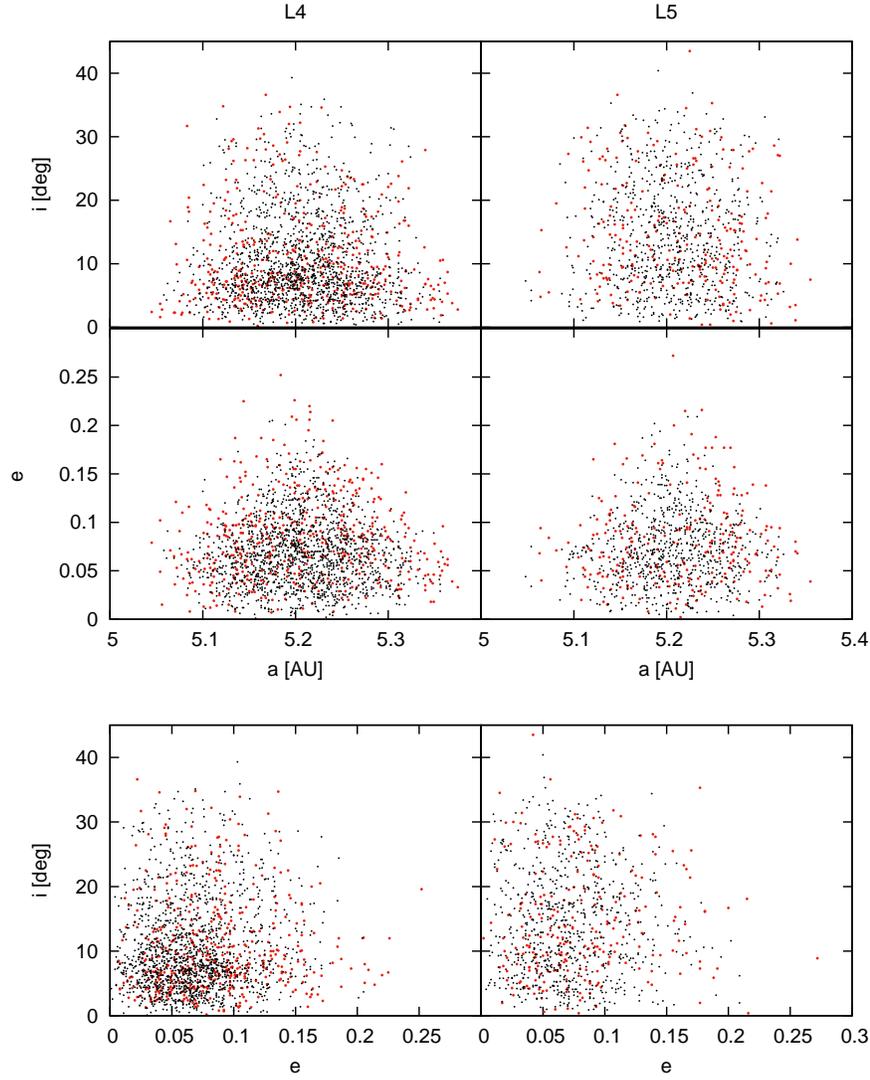}}
  \caption{Orbital elements of the numbered Trojans, i.e. the initial 
    conditions of the simulation. Black points represent the Trojan that are stable for 4.5 Gyr and the red 
    ones those that escape from the swarms.}  
\label{figci}
\end{figure}

The initial orbital elements of all $L_4$ and $L_5$ Trojans are shown in Fig. \ref{figci}. The black points represent 
the orbits of stable Trojans for 4.5 Gyr while escaped Trojans are represented in red.
  
Given the small step used in the integrations, the number of particles, and the long time interval, the 
initial orbits were divided in groups of nearly 20 Trojans. They were integrated by using several computers under 
the same conditions for several months. We performed a total of 48 runs for $L_5$ Trojans and 108 for $L_4$ ones.     

\section{General Results}
\label{res}

\subsection{Escape from $L_4$ and $L_5$}

We detect 466 (out of 1975) Trojans that escape from $L_4$, this is 23.6$\% $, and 250 (out of 997) 
Trojans that escape from $L_5$, i.e. 25.1 $\%$. Jupiter is the main cause for the escapes from both 
swarms, however, we have 1 $L_4$ Trojan that has its first encounter with Mars and 5 $L_4$ Trojans and 
3 $L_5$ Trojans with Saturn. 
 The analysis of the dynamical evolution of these particular Trojans that encounter other planet than Jupiter first, 
reveals that a slow diffusion among resonances is at work before the escape, as already noted by Robutel and Gabern  (2006). 
It is possible to see the typical behavior of objects going through secular and secondary resonances, 
 slowly increasing the eccentricity and changing the inclination, which eventually favors close encounters 
with the planets. This behavior is also found in some other Trojans that encounter Jupiter first.
However, the drastic change in the semimajor axis from which the escape of the resonance can be detected 
occurs after the encounter.

The number of escapees from $L_5$ is proportionally greater than 
that from $L_4$, in agreement with the results of D14, although the difference is smaller. 
We find that the escape rate from both Lagrangian points follow a linear trend with the time given by 
$a_{L_4} = 7.0398 \times 10^{-11} \pm 8 \times 10^{-14} $ and $a_{L_5} = 7.5590 \times 10^{-11} \pm 13 \times 10^{-14} $.

Following the same analysis as in D14, if the present unbiased asymmetry in the number of Trojans 
between $L_4$ and $L_5$ is $N_{s}(L_4)/N_{s}(L_5) = 1.4 \pm 0.2$ (Grav et al., 2011),
the original population of Trojans would have a primordial asymmetry of 
\begin{equation}
  N_0(L_4)/ N_0(L_5) = 1.373 \pm 0.204 .
  \label{ni}
\end{equation}
Therefore, the difference in the escape rate between $L_5$ and $L_4$, accounts for only 
$\sim 2 \%$ of the total asymmetry or, in other words, it has contributed to $\sim 7 \%$ of the present 
unbiased asymmetry.

\subsection{Post-escape}

The evolution of escaped Trojans was followed up to a collision with a planet or the Sun, or until a
heliocentric distance $r > 1000$ au (ejection was reached). From the 250 Trojans that escape from $L_5$, 2 of them end their 
evolution due to a collision with the Sun, 1 with Jupiter and the remaining objects are ejected. From the 466 
escaped $L_4$ Trojans, 16 of them collide with an object: 2 with the Sun, 2 with Saturn and 12 with Jupiter, 
and the remaining 450 are ejected. The different intrinsic rate of collisions between both escapees is remarkable; 
the proportional number of collisions by $L_4$ escapees is three times greater than that of $L_5$ escapees. 

 In Table \ref{tablaenc}, we show the fraction of escaped Trojans that have at least one encounter with each planet ($N_T$). 
From this set, we additionally calculate the quantity $N_e$, which is the fraction of encounters with each planet
with respect to the total number of encounters. For example, 466 Trojans escape from $L_4$, from which only 28 have 
at least one encounter with Venus, this is, $N_T = 6\%$. In addition, these 466 Trojans sum 229906 planetary encounters, 
from which 151 correspond to encounters with Venus, i.e. $N_e=0.07\%$. For $L_5$, the 250 Trojans that escape undergo 
131324 planetary encounters. We have found that most of the encounters 
occur with Jupiter and Saturn. Besides, the Trojans departing from $L_5$ have a larger $N_e$ for the inner 
Solar System including Jupiter than those departing from  $L_4$. The opposite is observed for the outer Solar System.

\begin{table}
  \caption{ Percentage of the escaped Trojans that have at least one encounter with each planet ($N_T$), 
  and percentage of encounters with each planet with respect to the total number of planetary encounters ($N_e$) 
  for $L_4$ and $L_5$ Lagrangian points.}
    \label{tablaenc}
  \centering
  \renewcommand{\footnoterule}{}  
  \begin{tabular}{lrrrrrr} \\
    \hline
    Planet  & \,& $L_4$ \,\,\,\,\,\,\,\,& \,\,\,\,\,\,\,\,\,\,\,\,\,\,\, &  $L_5$ \,\,\,\,\,\,\,\,\,  \\
    & $N_e (\%)$  &  $N_T (\%)$  &    $N_e (\%)$   &  $N_T (\%)$  \\ 
    \hline
    Venus     & 0.07  & 6   & 0.12  &  13    \\
    Earth     & 0.15  & 10  & 0.2   &  19   \\
    Mars      & 0.08  & 9   & 0.08  &  13  \\
    Jupiter   & 63.3  & 100 & 77.7  &  100   \\
    Saturn    & 21.9  & 97  & 17.2  &  95   \\
    Uranus    & 6.2   & 85  & 2.6   &  83   \\
    Neptune   & 8.3   & 76  & 2.1   &  76   \\
    
    \hline
  \end{tabular}
\end{table}

\begin{figure}[ht]
  \centering
  \begin{tabular}{cc}
    \includegraphics[width=0.95\textwidth]{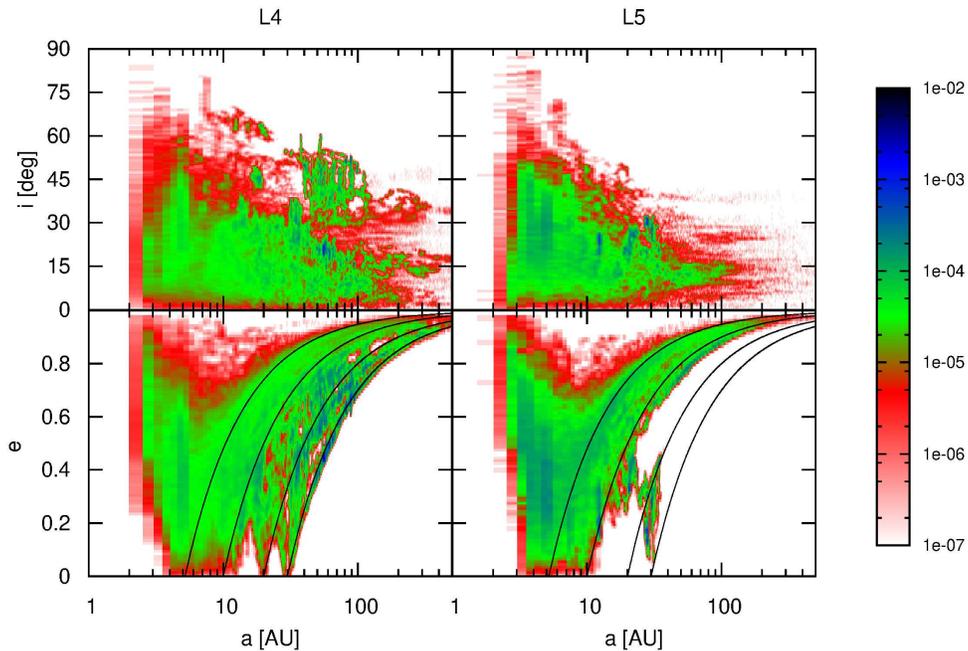}
  \end{tabular}
  \caption{Normalized time-weighted distribution of the dynamical evolution of escaped $L_4$ and $L_5$ Trojans 
    in the $(a,e)$ plane (left) and $(a,i)$ plane (right).   The color zones of these maps are regions
    with different degrees of probability where escaped Trojans can be found (blue for most visited regions, 
    red for least visited). The black curves correspond to constant perihelion values equal to the location of the giant planets.
    For a high resolution image, ask the authors.}
  \label{mapas}
\end{figure}

The whole evolution of escaped Trojans out of the swarms can be seen in Fig. \ref{mapas}. Those plots show the 
normalized time fraction spent by escaped Trojans in the orbital element space. The color code is indicative of the 
permanence time spent in each zone (blue for the most visited regions, red for the least visited). 
Then, those plots form dynamical maps of ``permanence'' in the different zones of the Solar System and
give a general  idea of the regions visited by escaped Trojans.

The regions of the Solar System occupied by escaped Trojans from $L_4$ and $L_5$ are similar but not equal. 
They cover similar ranges of orbital elements; however, differences in semimajor axes and inclinations can be 
observed in Fig. \ref{mapas}. $L_4$ escaped Trojans cover a wider range of semimajor axes than $L_5$ escapees.  
The inner regions of Jupiter's orbit are preferred by $L_5$ escaped Trojans; in Fig. \ref{mapas}, that zone 
is most visited by $L_5$ escapees than $L_4$ ones (i.e. there are blue strips in $L_5$ maps but not in $L_4$ maps). In this region, 
the densest zones correspond to the region near the $1:1$ mean motion, the Hilda region and the 
outer zone of the asteroid Main Belt. In the region outside Jupiter's orbit, $L_5$ escapees have perihelion 
distances near Saturn and there is a small structure near Neptune's perihelion whereas $L_4$ escapees cover almost all the 
external region with perihelion near all the giant planets. 
 There is also an island visited by $L_4$ escapees with a semimajor axes between 40 and 100
au, and an inclination between 30 and 60 degrees. This structure is generated by large variations in
inclination and eccentricity due to the Kozai mechanism inside
exterior MMRs with Neptune. We identified objects following this dynamic up to resonances $1 : 5$ with Neptune at
$a=88.07$ au and $1:6$ with Neptune at $a=99.45$ au.

\begin{figure}[ht]
  \centering
  \begin{tabular}{cc}
    \includegraphics[width=0.95\textwidth]{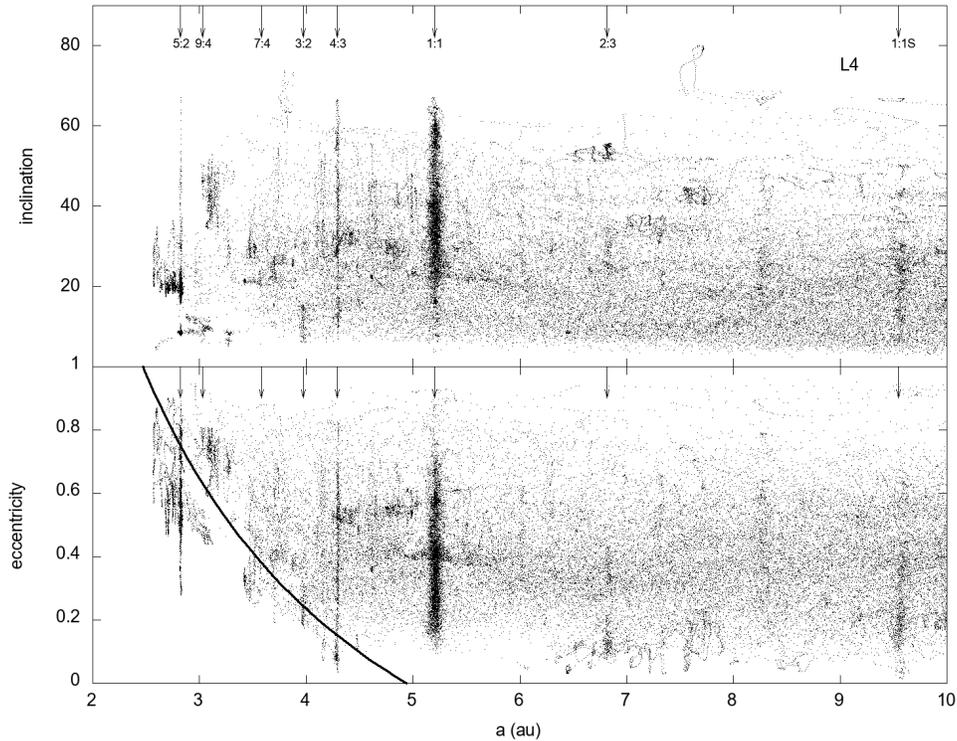}
  \end{tabular}
  \caption{Mean orbital elements of escaped Trojans from $L_4$. Small arrows indicate the 
    location of resonances. The points below the continuous curve correspond to orbital states completely inside 
    the orbit of Jupiter.}
  \label{aeil4}
\end{figure}

\begin{figure}[ht]
  \centering
  \begin{tabular}{cc}
    \includegraphics[width=0.95\textwidth]{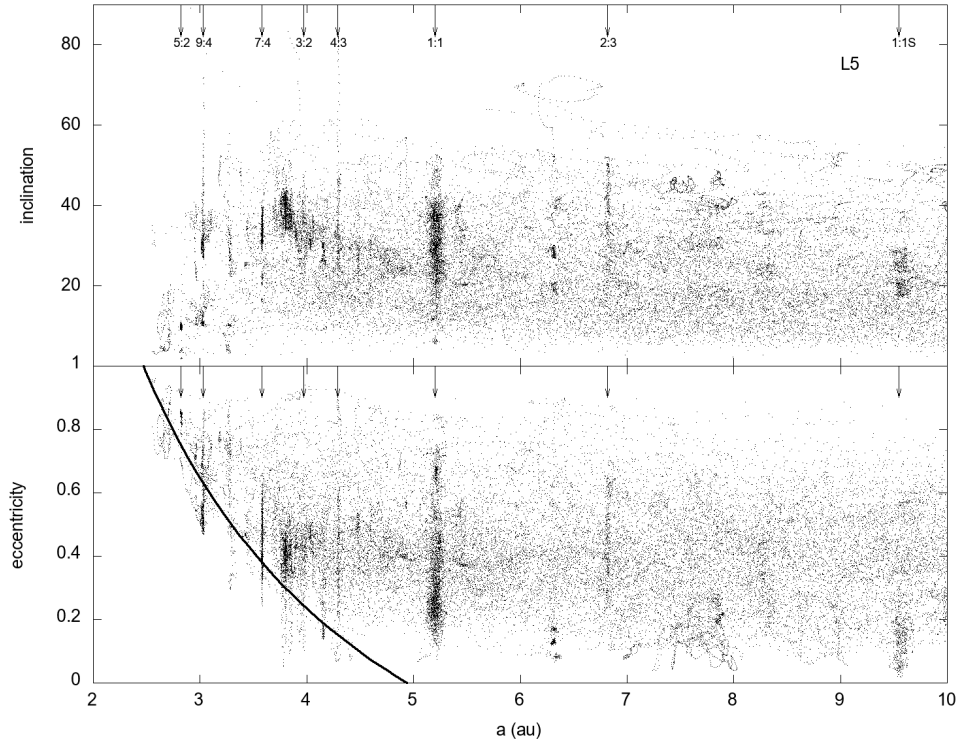}
  \end{tabular}
  \caption{The same as Fig. \ref{aeil4} for escaped Trojans from $L_5$.}
  \label{aeil5}
\end{figure}

To detect captures in MMR, we compute the mean orbital elements of the escaped Trojans by means of a running window 
of $10^4$ years every $10^3$ years, according to the following formula: $<E(t)> = 10^{-4} \int_{t-5000}^{t+5000} E(t')dt'$, 
where $E$ is the orbital element and $10^3$ years is equivalent to one orbital state. We compute 1644535 (173937) 
total orbital states calculated for the 466 (250) escaped Trojans from the $L_4$ ($L_5$) point. Figures \ref{aeil4} and \ref{aeil5} show 
the orbital states up to 10 au in the ($<a>,<e>$) and ($<a>,<i>$) planes. Several concentrations  for 
the mean semimajor axis around nominal values of MMRs are observed. For some resonances, the aphelion 
of the escaped Trojans decouples from Jupiter's orbit and evolves to an inner region far from the curve of constant 
aphelion with Jupiter. A similar behavior was found by Fern\'andez et al. (2018) for active Centaurs.
In particular, some escaped Trojans are temporarily captured in the exterior $2:3$ MMR  
with Jupiter and $1:1$ with Saturn. In Table \ref{tablares}, we show the number of orbital states in the most important resonances. 
We also calculate the number of escaped Trojans that remain in resonance for more than 20000 ($N_{20}$) and 
100000 ($N_{100}$) years. It is remarkable that the most populated resonances are first the co-orbital with Jupiter and 
then the co-orbital with Saturn.

\begin{figure}[ht]
  \centering
  \begin{tabular}{cc}
    \includegraphics[width=0.95\textwidth]{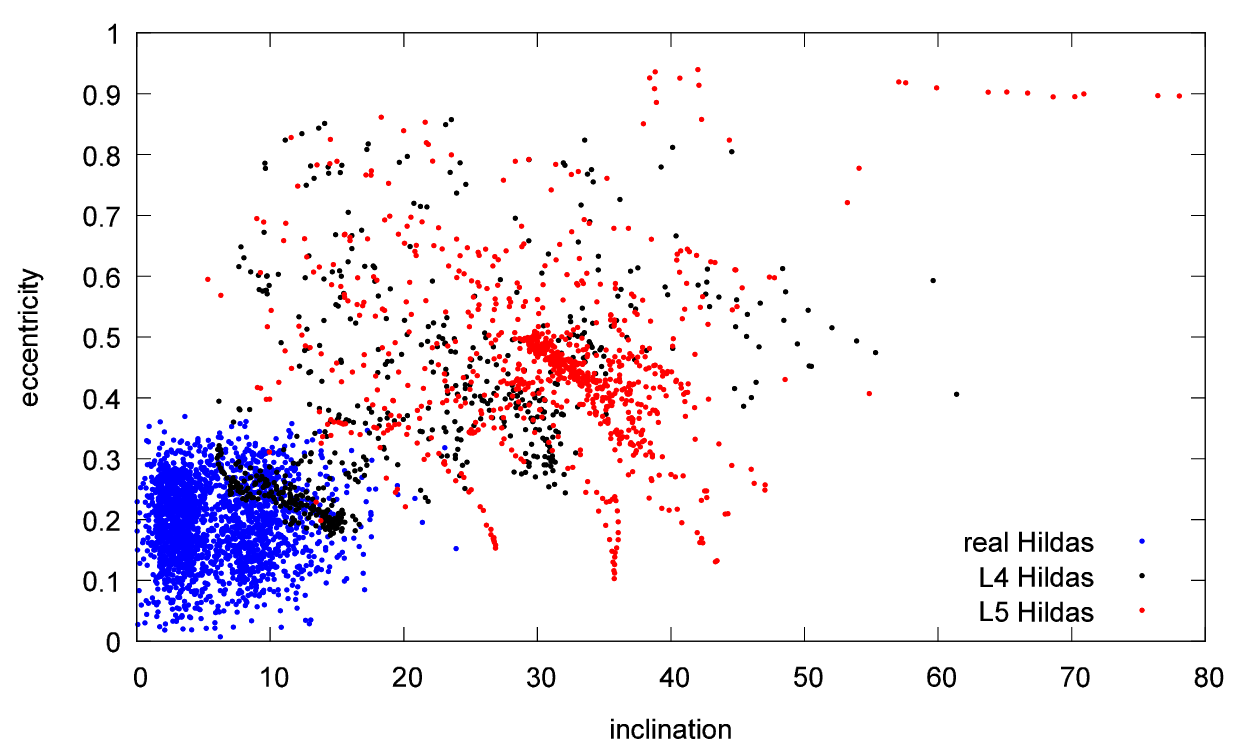}
  \end{tabular}
  \caption{Orbital elements for the Hildas and for the captured particles in the $3:2$ resonance.}
  \label{hildas}
\end{figure}

For the resonance $3:2$, we compared the orbital properties of the Hildas with the orbital parameters of the 
objects temporarily captured in the resonance $3:2$. Fig. \ref{hildas} shows the regions in the space  $(i,e)$ of the actual 
Hildas and the particles captured in the resonance $3:2$. The escaped Trojans evolve to the population of the Hildas but in 
general with very large eccentricity and inclination. Nevertheless, some orbital states are perfectly compatible with the actual Hildas.

\begin{table}
  \caption{The number of orbital states (see text) in the most important resonances, and the number of escaped Trojans that 
  remain in resonance for more than 20000 years ($N_{20}$) and 100000 years ($N_{100}$) for $L_4$ and $L_5$ 
  Lagrangian points.}
  \label{tablares}
  \centering
  \begin{tabular}{lr|rrr|rrr}
    \hline
    & & & escaped $L_4$ &  &  & escaped $L_5$ &  \\
    Resonance & $a$ [au] & states & $N_{20}$ & $N_{100}$ &  states & $N_{20}$ & $N_{100}$ \\
    \hline
    5:2 & 2.82 & 1543 & 5 & 3 & 144 & 2 & 1 \\
    9:4  & 3.03 & 464 & 5 & 2 & 796 & 7 & 2 \\
    7:4  & 3.58 & 358 & 11 & 1 & 787 & 11 & 1 \\
    3:2  & 3.97 & 510 & 18 & 1 & 983 & 14 & 2 \\
    4:3  & 4.29 & 936 & 15 & 3 & 340 & 7 & 1 \\
    1:1  & 5.20 & 6467 & 60 & 18 & 2535 & 46 & 8 \\
    2:3  & 6.82 & 516 & 8 & 1 & 413 & 8 & 1 \\
    1:1S  & 9.55 & 1594 & 48 & 1 & 852 & 19 & 2 \\
    \hline
  \end{tabular}
\end{table}

\begin{figure}[t!]
  \centerline{\includegraphics*[width=0.9\textwidth]{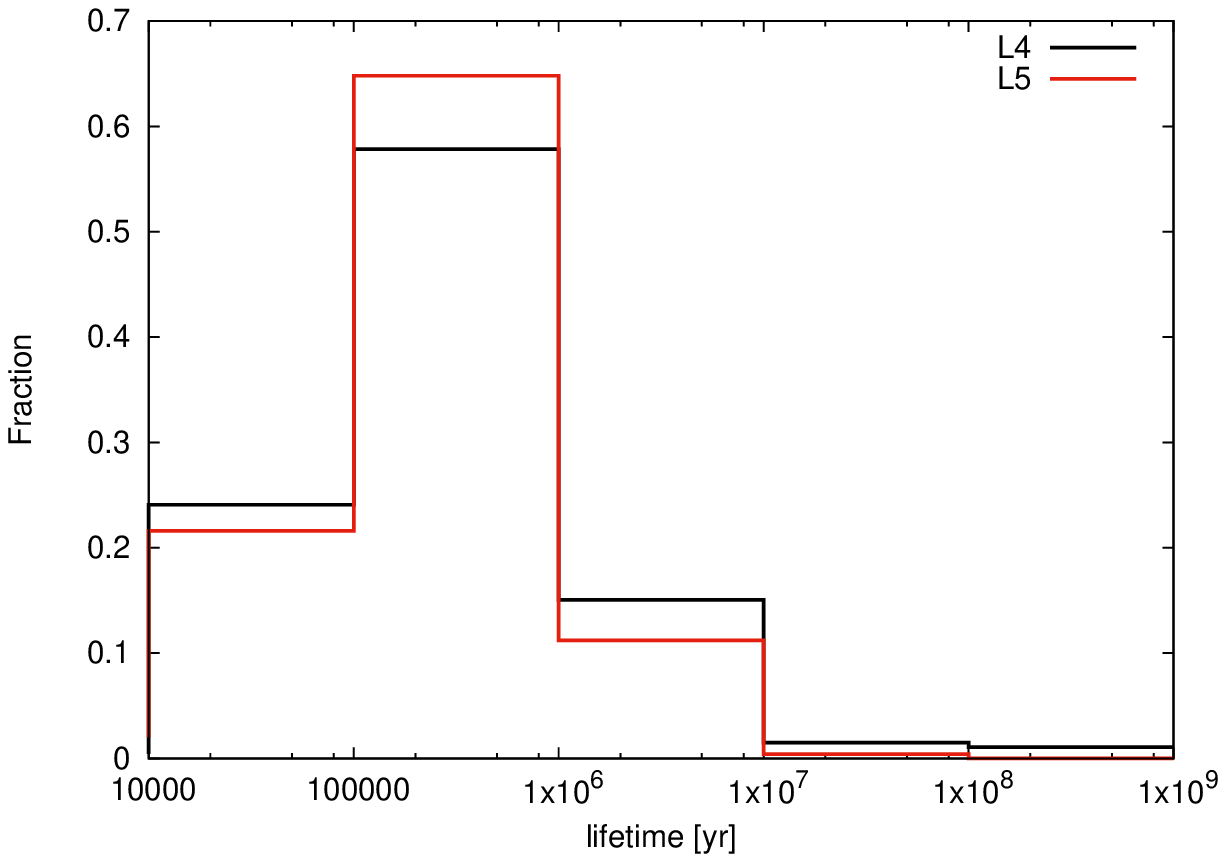}}
  \caption{Normalized distribution of mean lifetimes of $L_4$ (black) and $L_5$ (red) escaped Trojans.}
  \label{hlt}
\end{figure}

All the features observed in the orbital element distribution of escaped Trojans could be related to the 
different temporal evolution of both swarms. $L_5$ escapees have shorter lifetimes than $L_4$ ones as can be seen in 
Fig. \ref{hlt}, where the normalized distribution of lifetimes is plotted. 
We can also see that there are a few $L_4$ escapees that reach 
lifetimes greater than 100 Myr while there are no $L_5$ Trojans with these lifetime values.  

\begin{figure}[t!]
  \centerline{\includegraphics*[width=0.9\textwidth]{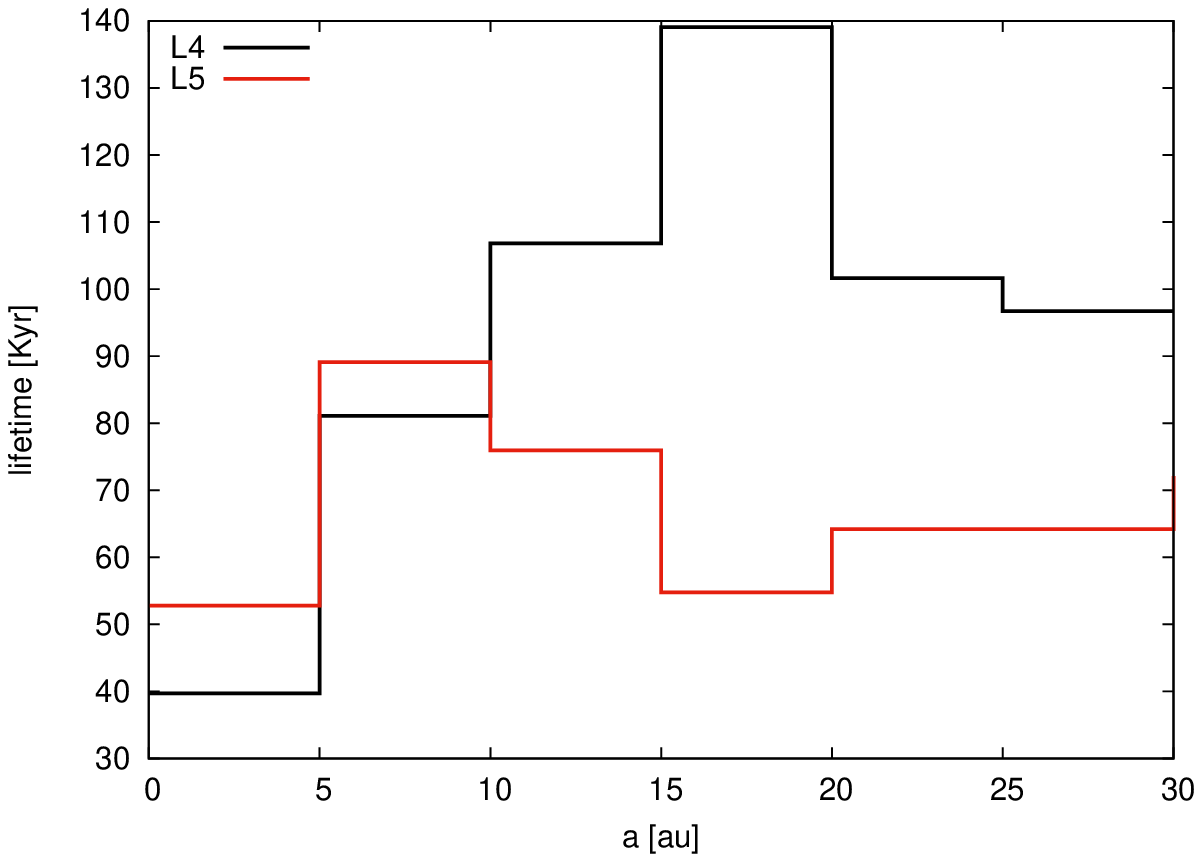}}
  \caption{ Mean lifetime of $L_4$ (black) and $L_5$ (red) escaped Trojans vs semimajor axis.}
  \label{halt}
\end{figure}

Another way of analysing  see the differences in the temporal evolution of escaped Trojans is shown in Fig. \ref{halt},
 where the mean lifetime versus the semimajor axis is plotted. We can see that for $a < 10$ au, the $L_5$ Trojans have
 a greater lifetime whereas this is reversed for $a > 10$ au and the difference grows up and become significant.  
In particular, for $15$ au$ < a < 20$ au, the difference is notable. This general behavior is intrinsic to the different 
evolution of $L_5$ and $L_4$ escapees, i.e. $L_5$  preferred the inner Solar System zones while $L_4$ escapees 
preferred the outer ones, as can be seen in Fig. \ref{mapas}.

The mean lifetime of escaped $L_5$ Trojans in the Solar System up to ejection or collision is 0.7 Myr while that of $L_4$ is 3.5 Myr, 
i.e. five times greater. However, the different behavior of $L_4$ and $L_5$ escapees is biased by a statistic of few objects. Then, 
to characterize the temporal evolution of escaped Trojans, we chose to evaluate the ``median lifetime'' as a typical 
lifetime of Trojans outside the resonance. In this case, we obtain that the median lifetime for $L_5$ escapees is $264000$ years, whereas 
that of $L_4$ escapees is $249000$, i.e. both of the same order. 

To test if the results for the escaped Trojans from $L_4$ and $L_5$ are representative of the real behavior of 
these populations, we take proportional samples of the initial populations,  extracting 20\% of the objects from 
each sample. For each subsample, composed by $80 \%$ of the original sample,  we performed all the calculations and statistics 
again. We repeated this procedure a few times. These experiments allowed us to better estimate the errors in our results. 
Since we obtain similar results for the different samples, we ca confirm that:
\begin{itemize}
\item The proportion of escaped Trojans from $L_5$ is  $25.1 \%$, which is slightly greater than that of $L_4$ of $23.6 \%$, 
  with an error of $0.2 \%$. 
\item The difference in the escape rate is also significant, though small, i.e. 
  $a_{L_4} = 7.04 \times 10^{-11} \pm 1 \times 10^{-12} $ and 
  $a_{L_5} = 7.56 \times 10^{-11} \pm 1 \times 10^{-13} $.  This implies a primordial asymmetry 
  of $  N_0(L_4)/ N_0(L_5) = 1.376 \pm 0.204 $.
\item The encounters with the planets follow the same trend in all experiments.
\item The mean lifetime of escaped $L_4$ Trojans is greater than that of $L_5$ due to the fact that 
 a few  $L_4$ Trojans have very long lifetime after escape.
\item The median lifetime of both escapees is of the same order.  
\end{itemize}

The different post-escape behaviour of $L_4$ and $L_5$  escapees can also be analytically investigated using the $\ddot{O}$pik theory (Opik,  1976; Valsecchi et al.,  2000). 
The Opik method analyzes an encounter of a particle with a planet and  gives a mean probability of collision with a 
planet and other parameters for an orbit with a given (a,e,i).  Then, following Valsecchi et al.  (2000) we consider the mean initial 
orbital elements of $L_4$ and $L_5$ escaped Trojans  and calculate the relative encounter velocity, the impact parameter, the probability of collision, 
the  mean time after which the object collides with the planet and the extreme changes in semimajor axis due to  the encounter. We consider planet Jupiter 
for the calculations.  Those values for both  Trojan swarms are shown in Table \ref{topik}.

\begin{table}
  \caption{$\ddot{O}$pik theory for escaped Trojans. The  input orbital elements (a, e, i) correspond to the mean initial 
orbital elements of $L_4$ and $L_5$ escaped Trojans. For those values, the relative encounter velocity (U), the impact parameter ($\sigma$),  
the probability of collision per orbital period ($P_{col}$), the mean time after which the object collides with Jupiter(t(col)) and the extreme changes in semimajor 
axis due to the encounter ($a_{max}$, $a_{min}$) are shown.}

\label{topik}
  \centering
  \begin{tabular}{l|c|c}
    \hline
    &  $L_4$  &  $L_5$   \\
    \hline
    a [au] & 5.2058 & 5.2118 \\
    e  & 0.0842 & 0.0813 \\
    i [degrees] & 10.6828 & 14.464 \\
    \hline
    U  & 0.204 & 0.264 \\
    $\sigma$ [$R_p$]  & 22.615 & 17.473 \\
    $P_{col}$ &  1.717$\times 10^{-5}$ & 1.019$\times 10^{-5}$\\
    t(col) [My]  & 0.692 & 1.167 \\
    $a_{max}$ [au] & 9.457& 12.958 \\
    $a_{min}$ [au] & 3.807 & 3.567 \\
    \hline
  \end{tabular}
\end{table}
 As can be seen, we obtain that the collision probability of $L_4$ escapees is  greater than for $L_5$ escapees. This is related to the different 
initial inclination of Trojans in boths swarms.  From $\ddot{O}$pik theory:
\begin{equation}
  P_{col} = \frac{U \sigma^2}{|U_x| \pi sin{i}},
  \label{pcol}
\end{equation}
where $U_x = \pm \sqrt {2 - 1/a - a(1-e^2) )}$ and $ [a] = a_j$.  
Then, since mean a and e are almost equal for both trojan swarms, the only variable that affects the result of an encounter is the different mean 
initial inclination. Orbits with greater initial inclinations give lower Tisserand constant ($T$) and then greater relative velocities ($U = \sqrt{3-T}$).
This is also transferred to the impact parameter $\sigma$ and all together to the probability of collision, as can be seen from Eq. {\ref{pcol}}. 
This is consistent with our numerical results of the rate of collision of escaped Trojans with the planets, especially with Jupiter, 
i.e., the proportional number of collisions by $L_4$ escapees is three times greater than that of $L_5$ escapees.
Another result from $\ddot{O}$pik theory shows that $L_5$ escapees have greater changes in semimajor axis than $L_4$ escapees and also 
higher speeds of encounter. In particular, $L_5$ Trojans go further than $L_4$ Trojans after an encounter. We think that this fact, 
together with $L_5$s'  higher speed of encounter, could make their evolution faster with respect to the $L_4$ ones. That is, $L_5$ Trojans 
would go further than the $L_4$ Trojans in each encounter and therefore spread out faster. Then, they would have proportionally 
fewer encounters with the planets beyond Jupiter than the $L_4$s (see Table \ref{tablaenc}) and their mean lifetime outside 
Jupiter would be lower than that of $L_4$ escapees in this zone. This could explain the shorter mean lifetimes of $L_5$ escapees with respect to 
the $L_4$ ones, found in our simulation. Also, this could explain the preference for $L_4$ escapees that  almost cover  all the 
external region with perihelion near all the giant planets, in contrast to $L_5$ escapees. Besides, the proportionally slower 
evolution of $L_4$ escapees through the external region would allow the existence of
very long-lived escaped Trojans as the few $L_4$ escapees that reach lifetimes greater than 100 Myr, while there are no $L_5$ 
Trojans with these lifetime values (see Fig. \ref{hlt}). We think that the application of $\ddot{O}$pik theory could help to explain those 
characteristics in the evolution  of escaped Trojans. However, it has to be regarded as an approximation since it does not take into 
account other perturbers than the planet considered.

\section{Contribution to other minor body populations}
\label{op}

To calculate the contribution of escaped Trojans to other minor body populations, we will define: 

\begin{itemize}
  
\item Comets: $q < 5.2$ au.
  
  \begin{itemize}
  \item JFCs:  $P < 20 $ yr and  $2 < T < 3.15$.

  \item no-JFCs:  $P > 20 $ yr or  $ T < 2$ or $ T > 3.15$.

  \end{itemize}
  
\item Centaurs:  $ q > 5.2$ au  and $ a < 30$ au.
  
\item TNOs: $a > 30$ au,

\end{itemize}
where $q$ is the perihelion distance, $a$ is the semimajor axis, $T$ is the Tisserand parameter with respect to Jupiter, 
and $P$ is the period.

We have considered the usual definition of JFCs but extended the limit of the Tisserand parameter to 3.15 taking into account 
that in fact, Jupiter's  orbit is slightly elliptic and then, orbits with $T$ values slightly above three allow 
close encounters with Jupiter (Di  Sisto et al., 2009). The above definitions were taken  in such a way that there is 
no overlap of populations. We have, according to perihelion distance, a comet population inside Jupiter's orbit, 
a Centaur population in the giant planet zone with semimajor axis inside Neptune's orbit, and  a TN population 
beyond Neptune. 
 
The contribution of escaped Trojans to each of the above defined populations is shown in Table \ref{t2}. We can see that 
almost all escaped Trojans from $L_4$ and $L_5$ reach the comet's zone and $\sim 90 \%$ go through the Centaur zone. The 
proportion of contribution from $L_4$ and $L_5$  is similar in those zones, though the mean lifetime of $L_4$ Trojans 
in a JFC zone is slightly smaller than that of $L_5$ Trojans, and the opposite occurs in the Centaur and TN zones. 
This last  topic is connected with what we have already mentioned about the 5 escaped $L_4$ Trojans that have long lifetimes 
in the Centaur and TN region. 

\begin{table}
  \caption{Percentages of escaped Trojans (with respect to the number of escapes) and mean lifetime ($\tau$) in 
  each population.}
  \label{t2}
  \centering
  \renewcommand{\footnoterule}{}  
  \begin{tabular}{lllll} \\
    \hline
    Population  & & $L_4$ \,\,\,\,\,\,& \,\,\,\,\,\,\,\,\,\, &  $L_5$ \,\,\,\,  \\
    & $N (\%)$   &  $\tau$ [yr]  &   $N (\%)$   &  $\tau$ [yr]    \\ 
    \hline
    JFCs     & 97  & 72300   & 97  &  88400    \\
    no-JFCs  & 93  & 87500 & 96  &  80900   \\
    Centaurs & 90  & 420000  & 84   &  272000   \\
    TNOs     & 78  & $3.9 \times 10^6 $  & 76  &  400000  \\
    
    \hline
  \end{tabular}
\end{table}

\begin{figure}[h!]
  \centerline{\includegraphics*[width=1.0\textwidth]{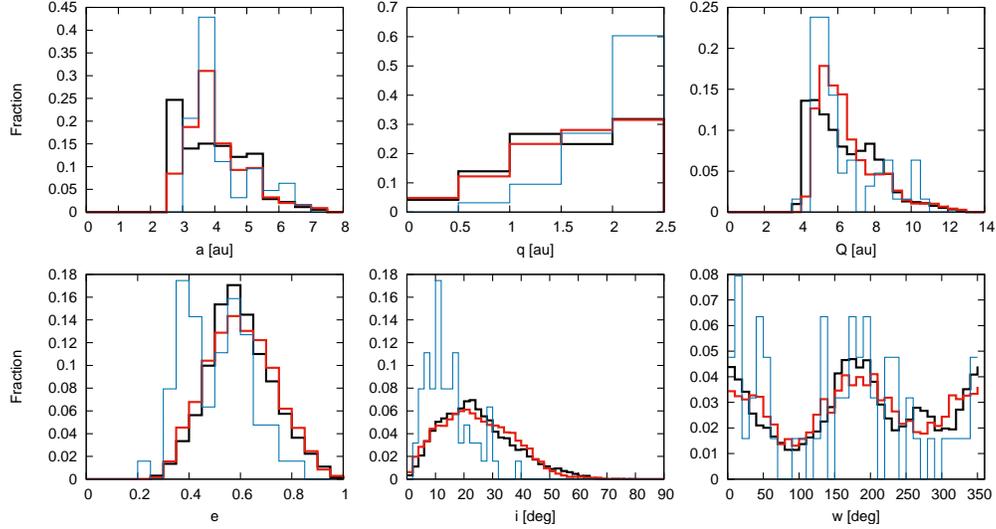}}
  \caption{Distribution of orbital states of escaped Trojans  
    ($L_4$ (black) and $L_5$ (red)) in the JFC zone with $q < 2.5$ au. The blue line corresponds to the observed 
    ``complete'' sample of JFCs ($q < 2.5 $ au and total absolute magnitude $H_T < 10$).}
  \label{hjfcq}
\end{figure}

\begin{figure}[h!]
  \centerline{\includegraphics*[width=1.0\textwidth]{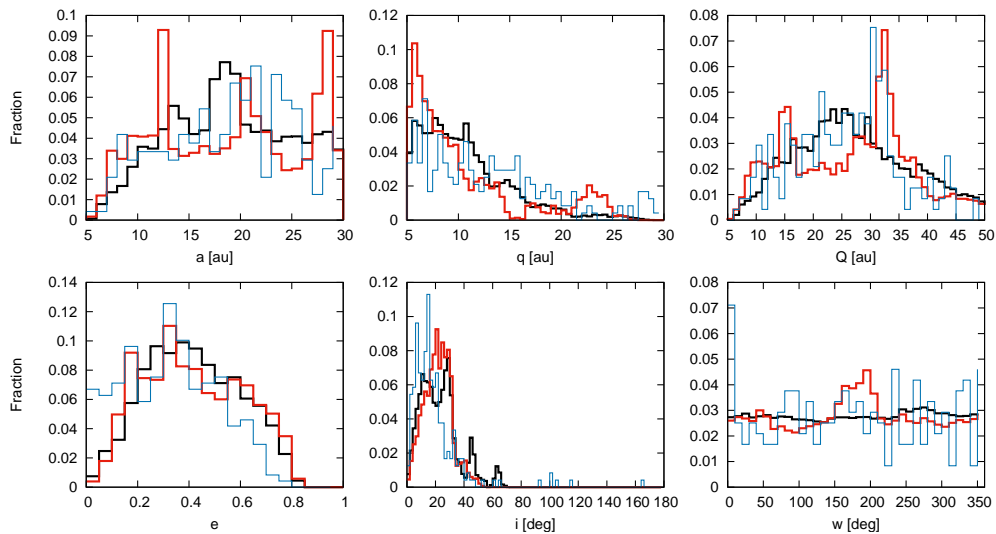}}
  \caption{Distribution of orbital states of escaped Trojans  
    ($L_4$ (black) and $L_5$ (red)) in the Centaur zone. The blue line corresponds to the observed Centaurs obtained 
    from the JPL Small-Body Database Search Engine. }
  \label{hcent}
\end{figure}

The orbital state distribution of escaped Trojans in the JFC and Centaur zones are plotted in Figs. \ref{hjfcq} 
and \ref{hcent}. The distribution of observed JFCs and Centaurs are also shown. The data were 
obtained from the JPL Small-Body Database Search Engine in September 2017.
 Fragments of disrupted comets were removed, leaving only one data point for each parent comet. To test the contribution of 
Trojans to JFCs, we consider the intrinsic real distribution of JFCs, i.e. a non-biased sample. We follow the 
reasoning proposed by Nesvorn{\'y} et al.  (2017) to select an unbiased sample of JFCs, and extract those JFCs with 
perihelion distances $q < 2.5$ au and absolute total magnitude $H_T < 10$. 
We have 63 known JFCs that satisfy these criteria. The comparison of the orbital element distribution of escaped 
Trojans and this ``complete'' sample is shown in Fig. \ref{hjfcq}. We can see that the semimajor axis and aphelion distances 
are reasonably well fitted, although there are differences. However, escaped Trojans reach smaller perihelion distances than JFCs. 
Eccentricities and inclinations of Trojans are higher than those of JFC ones and the argument of perihelion shows the
 same typical distribution of a population dominated by encounters with Jupiter as the JFC one.

In Fig. \ref{hcent} the orbital state distributions of escaped Trojans in the Centaur zone 
and known Centaurs are shown. We can see that escaped Trojans are compatible with observed Centaurs. In fact, the distribution 
of both spatial orbital elements and angular ones of escaped Trojans are similar to the observed distribution. Only 
Centaurs in low eccentricity and low inclination orbits are not compatible with escaped Trojans.  Two peaks are noticed 
in the inclination distribution of $L_4$ escaped Trojans near $40^{\circ}$  and $60^{\circ}$. They 
correspond to a few Trojans that remain for a long time on low eccentricity and high inclination  orbits, in some 
cases due to Kozai mechanism inside  MMRs. 

The previous analysis corresponds to our numerical simulation, which has aimed at evolving known Trojans. 
To evaluate the real contribution from Trojans, we have to take into account the real asymmetry between 
$L_4$ and $L_5$ as well as  the number of Trojans in each swarm given by Eq. (\ref{csd}). 
Then, the number of Trojans ejected out of the resonance per year will be given by:
\begin{equation}
  N_{ejec}(>R) = a_{L_i} N(>R) .
  \label{net}
\end{equation}

For example, there will be 5 $L_4$ Trojans and 4 $L_5$ Trojans  with radius $R > 1$ km ejected from the resonance every Myr. 
Or, for diameter $D > 1$ km,  2 $L_4$ Trojans are ejected every 100000 yrs and 3 $L_5$ Trojans every 200000 yrs.  

\begin{figure}[t!]
  \centerline{\includegraphics*[width=0.9\textwidth]{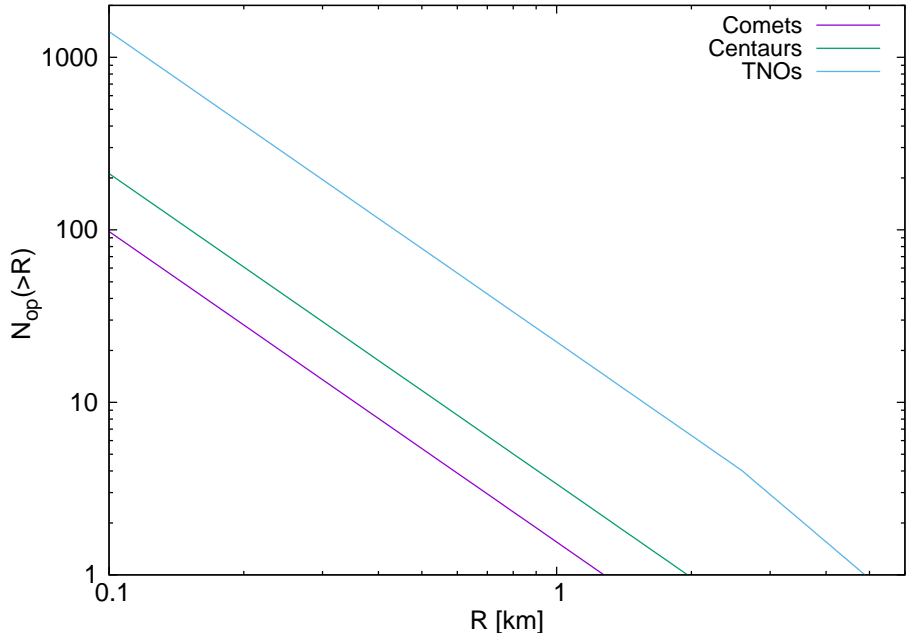}}
  \caption{Cumulative number of escaped Trojans in each minor body population.}
  \label{fignop}
\end{figure}

The number of escaped Trojans in each minor body population would be given by:
\begin{equation}
  N_{op} (>R)= a_{L_i} \, N (>R) \,\, \tau,
  \label{nop}
\end{equation}
and it is plotted in Fig. \ref{fignop}. We have added the contribution from both swarms and also the JFCs and 
no-JFCs into comets in the plot.  However, we have noticed that considering the real asymmetry and the real number of 
Trojans given by Eq. (\ref{csd}),  the contribution to JFCs and no-JFCs from $L_4$ and $L_5$ is almost the same, but the 
$L_4$ contribution to Centaurs is two times greater than that of $L_5$, and the $L_4$ contribution to TNOs is $\sim 15$ times 
that of $L_5$. This difference, however, is based mostly in the five $L_4$ Trojans that have long lifetime in the 
Centaur and TNO regions and then it should be taken with caution because of a low number statistic. 
Beyond this, the contribution of escaped Trojans to Comets, Centaurs and TNOs is negligible. In fact, for example, 
Di  Sisto et al. (2009) estimate that there would be $\sim 450$ JFCs with $R > 1$ km within Jupiter's orbit, and the 
contribution of Trojans would be only 1. In the case of Centaurs and TNOs, the number of Trojans with $R > 1$ km 
would be 3 and 20 respectively, which is orders of magnitude lower than the total number of both populations. 

Nevertheless, small Trojans in cometary orbits could resemble Asteroids in Cometary Orbits (ACOs). For example, we have 
$\sim 6$ Trojans in cometary orbits with $D > 1$ km. Tancredi (2014) identified 203 ACOs belonging to the 
Jupiter-family group, thus, the contribution from Trojans would be minor. 
We have also tested whether there are Trojans in the NEO population ($q < 1.3$ au), and have we obtained that this contribution 
is also negligible. We could expect 2 escaped Trojans with $D > 1$ km in the NEO population with dynamical lifetimes of $\sim 30000$ yrs. 
Almost all of those escaped Trojans with $q<1.3$ au are in a JFC-orbit. Fern{\'a}ndez and Sosa (2015) carried out orbital 
integrations of JFCs in NEO orbits and obtained that 
there is a fraction of them that show stable asteroidal orbits with lifetimes greater than 10000 yrs. They obtained that  
 at least 8 JFCs in NEO orbits show this stable behavior and being them km- and sub-km size bodies, they attribute 
their long lifetime to a mostly rocky composition and  might have a source region in the outer main asteroid belt. 
So, a fraction of those objects could come from the Trojan swarms.     

It is interesting that a fraction of escaped Trojans go through the zone of Encke type comets (ETC), defined as those with 
$T > 3$ and $a < a_{Jup}$.  We could expect 1 ETC with $D > 1$ km in this zone. However, the exact orbital elements of 
2P/Encke are not reached by escaped Trojans. 

Another interesting case is the Shoemaker Levy 9 (SL9) impact with Jupiter. In our simulations we have a great number of collisions 
of escaped Trojans with Jupiter. Also, the number of close encounters with Jupiter within the Roche limit is important: we have 
5 $L_5$ escaped Trojans and 17 $L_4$ escaped Trojans that encounter Jupiter within its Roche limit. Within this radial distance to the planet, 
an object could be fragmented and end up impacting with the planet, as did the SL9. The rate of those encounters is, in fact, 
constant and then we could expect, for example, 5 ``SL9 case''- Trojans with $D > 1$ km from $L_4$ and 2 from $L_5$ every 10 Myr, or in total 
one escaped Trojan with $D > 1$ km every 1.4 Myr would cross the Roche limit of Jupiter and then it would fragment and end up 
impacting Jupiter. Di  Sisto et al. (2005) found that one escaped Hilda asteroid with $D > 1$ km would impact Jupiter
every 65000 yrs, being this rate of collision much greater than our estimated rate for escaped Trojans. 

\section{Discussion and Conclusions}

Trojan asteroids are located in stable reservoirs and have long dynamical lifetimes. However, some 
of them are located in unstable zones and then are capable to escape from the swarms. In this paper, we have analyzed  
the observed $L_4$ and $L_5$ Trojan population trough numerical simulations in order to study the dynamical behavior of the escaped 
Trojans and their contribution to other minor body populations. 
We obtain that the number of Trojans that escape from $L_5$ in the age of the Solar System is proportionally greater than 
that from $L_4$. The difference is small, we have 25.1 $\pm$ 0.2$\%$ of Trojans from $L_5$ and  23.6 $\pm$ 0.2$\% $ from $L_4$. 
The escape rate from both swarms along the integration time can be fitted by a linear relation and the one from $L_5$ 
 is greater than that from $L_4$ over time.  Those results are qualitatively similar to D14.

The dynamical evolution of escaped Trojans was studied up to their ejection of the Solar System or collision with the Sun or a 
planet.  The main general results of the simulation are the following:

\begin{itemize}
 \item The proportional number of collisions by $L_4$ escapees is three times greater than that of $L_5$ escapees, being  
the great majority of them with Jupiter although few collisions with the Sun and Saturn are also registered. 
This is a consequence of the different mean inclination of Trojans in both swarms, as we have demonstrated using $\ddot{O}$pik theory.
\item Most of the encounters are with Jupiter, but there are also with other planets. We found a greater 
relative proportion of encounters by $L_5$ Trojans with Jupiter and the inner planets with respect to $L_4$ Trojans, 
and the reverse is observed for the outer planets (Saturn, Uranus and Neptune). 
\item $L_4$ and $L_5$ escaped Trojans cover similar regions of orbital elements; however, $L_5$ escapees preferred 
the  regions interior to Jupiter's orbit and $L_4$ escapees cover almost all the external region with perihelion 
near all the giant planets.
\item  $L_5$ Trojans spent 0.7 Myr up to ejection or collision  whereas $L_4$ spent 3.5 Myr. 
But this difference is mainly due to five $L_4$ Trojans that have long dynamical lifetime
 (i.e. $>100$ Myr) once they escape, slowly evolving in the Centaur and TNO regions. A better characterization of  
the temporal evolution of escaped Trojans is then the ``median lifetime'' that is of the same order for both escapees,
 i.e. $264000$ yrs for $L_4$ and $249000$ yrs for $L_5$ escapees. 
\end{itemize}

 The contribution of Trojans to other minor body populations was analyzed. We found that  
almost all escaped Trojans from $L_4$ and $L_5$ reach the comet's zone,  $\sim 90 \%$ go through the Centaur zone 
and only $L_4$ Trojans reach the transneptunian zone. In particular, we note that the distribution 
of both spatial orbital elements and angular ones of escaped Trojans are similar to the observed Centaur 
orbital element distribution. Considering the real asymmetry between $L_4$ and $L_5$ and the number 
of Trojans in each swarm given by Eq. (\ref{csd}), we obtained  the number of Trojans ejected out of the resonance 
per year. Then, for example, there are 5 $L_4$ Trojans and 4 $L_5$ Trojans  with radius $R > 1$ km ejected from the 
resonance every Myr. Or, for diameter $D > 1$ km,  18 $L_4$ Trojans are ejected every Myr and 14 $L_5$ Trojans every Myr.  
The results of the present paper are based on the dynamical evolution 
of Trojans, and then, the escape rate from the swarms is due only to dynamical instabilities on the resonance. However, 
the collisional evolution of Trojans could be responsible for the escape of some of them, too. In fact, de El\'{\i}a and Brunini (2007)
analyze the collisional evolution of $L_4$ Jovian Trojans. They obtained that most of the bodies ejected
from the $L_4$ swarm are small; one could expect up to $\sim $ 50 Trojans ejected from $L_4$ swarm 
with $D > 1$ km per Myr. So, the escape rate by collisions would be roughly three times the dynamical rate of escape at 
least for small bodies, since large Trojans are unaffected by collisional evolution (de El\'{\i}a and Brunini, 2007). 
Then, the real number of escapes of small Trojans would be the sum of the collisional and dynamical removal.

 We calculated the number of escaped Trojans in each minor body population. Considering the real asymmetry and the 
real number of Trojans,  the contribution to JFCs and no-JFCs from $L_4$ and $L_5$ is almost the same, but the $L_4$ 
contribution to Centaurs and TNOs is orders of magnitude greater than that of $L_5$. 
Considering the collisional removal, and assuming that Trojans that escape by collisions follow the same dynamical 
behavior that the ones removed by dynamics, we would have a minor contribution of Trojans to comets and Centaurs. 
For example, from our dynamical simulation,  we could expect $\sim 20$ Trojans in cometary 
orbits with $D > 1$ km or $\sim$ 7 in JFC-orbit. 

 There are some specific regions where escaped Trojans could be important if considering dynamical plus collisional removal.  
\begin{itemize}
 \item We could expect 8 escaped Trojans with $D > 1$ km in an NEO-JFC orbit with dynamical lifetimes of $\sim 30000$ yrs. 
  Fern{\'a}ndez and Sosa (2015) obtained that at least 8 JFCs in NEO orbits show a stable behavior and their composition is 
  mostly rocky; their source region is in the outer main asteroid belt. 
  \item A fraction of escaped Trojans go through the zone of Encke type comets. We could expect 4 
ETC with $D > 1$ km in this zone. However the exact orbital elements of 2P/Encke are not reached by escaped Trojans.
\item We could expect that 1 escaped Trojan with $D > 1$ km every  350000 yrs would cross the Roche limit of Jupiter, then 
fragment and impact Jupiter as the SL9 case. This estimation is already smaller than the contribution of Hildas to this 
type of objects, but it is not negligible. 
\end{itemize}

Although the contribution of escaped Trojans to other minor body populations would be minor, at least it could explain 
some peculiarities observed in some populations of our Solar System.

\vspace*{0.5cm}
\noindent{\bf Acknowledgments:}
The authors wish to express their gratitude to the FCAGLP for extensive use of their computing facilities and acknowledge 
the financial support by IALP, CONICET and Agencia de Promoci\'on Cient\'{\i}fica, through the grants PIP 0436  and
 PICT 2014-1292.  We would like to thank Gabinete de Ingl\'es de la Facultad de Ciencias Astron\'omicas y Geof\'{\i}sicas 
 de la UNLP for a careful language revision. We also acknowledge the constructive comments from Christos Efthymiopoulos and an 
 anonymous referee which helped us improve the article.

\vspace*{0.5cm}

\end{document}